\documentclass[printer]{aa}
\usepackage{times}
\usepackage{natbib}
\usepackage[dvips]{graphicx}
\usepackage{epsfig,longtable,lscape}
\usepackage{epsfig}
\usepackage{rotating}
\bibpunct[]{(}{)}{;}{a}{}{,}

\def\ltsima{$\; \buildrel < \over \sim \;$}
\def\lsim{\lower.5ex\hbox{\ltsima}}
\def\gtsima{$\; \buildrel > \over \sim \;$}
\def\gsim{\lower.5ex\hbox{\gtsima}}
\def\approxgt{\mathrel{\hbox{\rlap{\lower.55ex \hbox {$\sim$}}
        \kern-.3em \raise.4ex \hbox{$>$}}}}
\def\approxlt{\mathrel{\hbox{\rlap{\lower.55ex \hbox {$\sim$}}
        \kern-.3em \raise.4ex \hbox{$<$}}}}

\def\Chandra{{\em Chandra}}
\def\XMM{{\em XMM-Newton}}

\def\BDone{{BD+37$^\circ$ 442}}
\def\BDtwo{{BD+37$^\circ$ 1977}}
\def\BDthree{{BD+28$^\circ$ 4211}}
\def\HD{HD 49798}

\begin{document}

\title{Three new X-ray emitting sdO stars discovered with \Chandra}
\author{N. La Palombara\inst{1}, P. Esposito\inst{1}, S. Mereghetti\inst{1}, A. Tiengo\inst{1,2,3}}

\institute{INAF, Istituto di Astrofisica Spaziale e Fisica Cosmica - Milano, via Bassini 15, I-20133, Milano, Italy
\and IUSS-Istituto Universitario di Studi Superiori, piazza della Vittoria 15, I-27100 Pavia, Italy
\and Istituto Nazionale di Fisica Nucleare, Sezione di Pavia, via A. Bassi 6, I-27100 Pavia, Italy}

\titlerunning{\Chandra\ observations of bright sdO stars}

\authorrunning{La Palombara et al.}

\abstract{\textit{Context.} X-ray observations of sdO stars are a useful tool to investigate their properties, but so far only two sdO stars were detected at X-rays.\\
\textit{Aims.} We aimed to perform the first systematic search for X-ray emission from sdO stars to characterize the X-ray emission from single and binary sdO stars.\\
\textit{Method.} We observed a complete flux-limited sample of 19 sdO stars with the \Chandra\ HRC-I camera to measure the count rate of the detected sources or to set a tight upper limit on it for the undetected sources.\\
\textit{Results.} We obtained a robust detection of \BDtwo\ and Feige 34 and a marginal detection of \BDthree. The estimated luminosity of \BDtwo\ is above $10^{31}$ erg s$^{-1}$, which is high enough to suggest the possible presence of an accreting compact companion. This possibility is unlikely for all the other targets (both detected and undetected), since in their case $L_{\rm X} \lsim 10^{30}$ erg s$^{-1}$. On the other hand, for all 19 targets the estimated value of $L_{\rm X}$ (or its upper limit) implies an X-ray/bolometric flux ratio that agrees with log($L_{\rm X}/L_{\rm bol}$) = --6.7 $\pm$ 0.5, which is the range of values typical of main-sequence and giant O stars.\\
\textit{Conclusions.} The observing campaign performed by \Chandra\ has discovered three new X-ray emitting sdO stars; for one of them the observed X-ray flux might be emitted by an accreting compact companion, while for the other two stars it is most probably due to intrinsic emission. The same is possibile for the 16 undetected stars.\\
}

\maketitle

\section{Introduction}\label{sec:1}

Hot subdwarf (sd) stars are blue low-mass stars, commonly located at high Galactic latitudes, which are at an advanced evolutionary phase; they have already lost most of their hydrogen envelope and are now burning their helium core \citep[see ][ for a review]{Heber09}. Based on their spectral properties, hot sd stars are classified into two different types: the cooler B-type subdwarfs (sdB), which have T$_{\rm eff}<$ 40,000 K and usually display no or weak helium lines in their spectra, and the hotter O-type subdwarfs (sdO), which have T$_{\rm eff}>$ 40,000 K and, in most cases, are helium rich \citep{Hirsch+08}. While the sdB stars form a rather homogeneous group, the class of sdO stars is very heterogeneus: they cover a wide range of effective temperatures (T$_{\rm eff}$ = 40--100 kK), surface gravities (log($g$) = 4--6.5), and helium abundances \citep{HeberJeffery92,Heber+06}. Therefore, sdO stars can be subdivided into He-poor and He-rich, according to the atmospheric abundances, and into `luminous' and `compact', depending on their low or high values of log($g$), respectively \citep{Napiwotzki08b}.

The two classes of sd stars are very different also with respect to their origin. The sdB stars belong to the extreme horizontal branch (EHB) stars \citep{Heber86}: unlike normal HB stars, they evolve directly to the white-dwarf cooling sequence after the core helium exhaustion, without ascending the asymptotic giant branch (AGB), since their hydrogen envelope is too thin to sustain hydrogen burning. The origin of sdO stars is more complex. The luminous He-poor stars are low-mass post-AGB stars, while the compact ones are post-EHB stars and very probably are descendant of sdB stars. It is more difficult to explain the origin of the He-rich sdO stars. The luminous ones are on the post-AGB track, while the compact ones cannot evolve from sdB stars: in their case the most probable formation mechanism is either the merging of two He-core or C/O-core white dwarfs \citep{Iben90,SaioJeffery00,SaioJeffery02} or the so-called \textit{late hot-flasher} scenario \citep{Brown+01}.

Radial velocity surveys of sd stars have shown that a high percentage ($>$ 40 \%) of the cool sdB stars occur in close binary systems \citep{Maxted+01,Napiwotzki+04, Morales-Rueda+06}, while the binary fraction of sdO stars is much lower \citep{Napiwotzki08a}: therefore, binary evolution can play an important role in the formation of subdwarf stars, particularly in that of sdBs. Three mechanisms have been identified to form subdwarf stars, starting from systems with stellar components \citep{Han+02,Han+03}: 1) one or two phases of common-envelope and spiral-in evolution; 2) one or two phases of stable Roche-lobe overflow; and (3) the merger of two helium-core white dwarfs. Moreover, a common-envelope phase involving a substellar companion has been suggested as a possible formation scenario \citep{Soker98,Charpinet+11}. In some cases the final outcome is the formation of a binary system with a subdwarf star and a compact companion; usually, the compact companion of the hot subdwarf is expected to be a white dwarf (WD), but if the mass of the original stars is large enough, a neutron star (NS) or black hole (BH) companion can be formed.

The observation of hot subdwarfs at X-ray wavelengths can be a very useful tool to investigate their properties. First of all, they can be characterized by intrinsic X-ray emission. In fact, main-sequence, giant and supergiant early-type (O and B) stars are a well-known class of soft X-ray sources, with $L_{\rm X} \sim 10^{31}-10^{32}$ erg s$^{-1}$. Their X-ray emission is attributed to turbulence and shocks in their strong winds \citep{LucyWhite80}, and the X-ray and the bolometric luminosities are linked by the canonical relation $L_{\rm X} \sim 10^{-7} \times L_{\rm bol}$ \citep{Pallavicini+81,Sciortino+90,GuedelNaze09}. Although the hot subdwarfs are characterized by lower luminosities (log($L_{\rm bol}/L_{\odot}$) $\lsim$ 4 instead of 5--6), they can have winds with mass losses of up to 10$^{-8}$ and 10$^{-10}$ M$_{\odot}$ y$^{-1}$ for sdO and sdB stars, respectively \citep{Hamann10,JefferyHamann10}. Therefore, hot sd stars might also be X-ray emitters of the same type, and it is interesting to investigate whether the above average relation extends to such low luminosities as well. In addition, the observation of X-ray emission could indicate the presence of a compact companion that accretes matter from the subdwarf wind: in this case the measured X-ray luminosity can provide useful information on the binary orbit and the mass-loss rate from the subdwarf star.

A Swift/XRT search for X-ray emission from candidate sdB+WD/NS binaries, selected from optical spectroscopy and photometry \citep{Geier+10}, gave negative results \citep{Mereghetti+11}, probably because of the weak winds of sdB stars, which are unable to provide a high enough accretion rate. On the other hand, the only two sdO stars for which pointed X-ray observations are available both showed a clear X-ray emission.

The first case is \HD, which is the brightest (V = 8) sdO star and is known since 1970 as a single-lined spectroscopic binary \citep{Thackeray70,SticklandLloyd94}. The compact nature of its companion, undetected in the optical/UV, was indicated by the presence of soft X-ray emission with a periodicity of 13.18 s \citep{Israel+97}. A long \XMM\ observation allowed us to constrain the orbital parameters and the X-ray spectrum and luminosity ($L_{\rm X} \simeq 3\times10^{31}$($d$/650 pc)$^{2}$ erg s$^{-1}$) and to establish that the most likely companion is a WD \citep{Mereghetti+09}. The X-ray flux from \HD\ does not disappear completely when the X-ray pulsar is eclipsed by the sdO star. The spectrum observed during the eclipse shows emission lines of H- and He-like nitrogen, an overabundant element in \HD. This emission could result from reprocessing of the accretion-powered X-rays in the sdO wind, but also from \HD\ iteself \citep{Mereghetti+13}. Its value of $L_{\rm X}/L_{\rm bol} \sim 10^{-7}$ agrees with the average value found in main-sequence early-type stars.

The second sdO star detected in X-rays is the bright He-rich star \BDone, which has stellar parameters very similar to those of \HD. Like \HD, it shows P-Cygni UV line profiles indicating wind mass-loss at a rate of $\sim3\times10^{-9}$ $M_{\odot}$ yr$^{-1}$ \citep{JefferyHamann10}, but no evidence of binary nature was reported from optical/UV photometry and spectroscopy. We recently observed it with \textit{XMM-Newton} and discovered soft X-ray emission, with a spectrum similar to that of \HD, and pulsations at 19.16 s, indicating that \BDone\ is also a binary with a NS or WD companion \citep{LaPalombara+12}. The luminosity is in the range $10^{32}-10^{35}$ erg s$^{-1}$ (for a source distance of 2 kpc), consistent with wind accretion onto a WD or, more likely, onto a NS. However, optical observations show no variations in the radial velocity of \BDone\ and, hence, there is no dynamical evidence for the existence of a compact companion yet \citep{Heber+14}.

Prompted by our results on \HD\ and \BDone, we planned a survey with \Chandra\ HRC-I of a complete flux-limited sample of sdO stars. Our aim was to perform the first systematic search for X-ray emission from this type of stars to characterize the X-ray emission from single and binary sdO stars. In \S\ref{sec:2} we present the sample of the selected sources and the adopted observing strategy; in \S\ref{sec:3} we describe the observations, the data reduction, and the results; finally, in \S\ref{sec:4} we briefly discuss these results and compare them with those obtained for \HD\ and \BDone.

\section{Source sample and observing strategy}\label{sec:2}

In Table~\ref{targets} we report our sample of selected sources, which consists of all the subdwarfs of O spectral type brighter than V = 12. They are characterized by high effective temperatures (T$_{\rm eff} \gsim$ 40,000 K) and by a wide range of surface gravity (log($g$) = 4--6.5); therefore, the selected sample includes both luminous (low-gravity) and compact (high-gravity) stars. For comparison, we also report in Table~\ref{targets} the parameters of \HD\ and \BDone. For each source we estimated the bolometric correction using the relation BC = 27.66 - 6.84$\times$log(T$_{\rm eff}$), which holds for O-type stars \citep{Vacca+96}; then we used it to obtain the bolometric flux and luminosity. However, the temperature estimates of the hot subdwarf stars can be affected by substantial uncertainties. This is well known for sdB stars \citep[e.g.][]{Green+11} and is proving to be the case also for sdO stars. For example, for \BDone\ \citet{Heber+14} reported T$_{\rm eff} \simeq$ 56 kK instead of the 48 kK estimated by \citet{JefferyHamann10}. At any rate, this temperature difference implies a variation of less than 0.5 in the bolometric correction and, hence, of 50 \% at most in the bolometric luminosity. None of the selected sources was previously observed with sensitive X-ray telescopes. They were not detected in the ROSAT All Sky Survey (RASS), which has a typical sensitivity of a few $10^{-13}$ erg cm$^{-2}$ s$^{-1}$ assuming a soft spectrum (e.g. a blackbody with kT = 50 eV)\footnote{For comparison, \HD\ (which has $f_{\rm X} \simeq 5\times 10^{-13}$ erg cm$^{-2}$ s$^{-1}$ out of the eclipse) was clearly detected in the RASS, while \BDone, which is about one order of magnitude fainter, was not.}. All but one of our targets are closer than 1 kpc, which implies a low interstellar absorption; considering the expected soft spectrum, this is an important advantage for their detection in X rays. The only exception is \BDtwo, which stands out among the other sources: it is located much farther away than the other stars (d $\simeq$ 2.6 kpc) and its intrinsic luminosity is much higher (log($L_{\rm bol}/L_{\odot}$) = 4.37). We note that this source is spectroscopically very similar to \BDone\ \citep{JefferyHamann10}, therefore it is very interesting to investigate whether it is characterized also by a similar X-ray emission as well.

We decided to observe our targets with the \textit{Chandra} HRC-I camera, since it has high efficiency at low energies and excellent spatial resolution, which implies a negligible background level in the small source-extraction region. Therefore, it provides the highest signal-to-noise ratio for the soft spectrum expected for these sources; in addition, the high spatial resolution makes it possible to minimize the source confusion.

\section{Observations, data reduction and results}\label{sec:3}

The sources reported in Table~\ref{targets} were observed with \textit{Chandra} HRC-I between March and October 2013; in Table~\ref{observations} we report the main parameters of the observations. In most cases the effective exposure time was about 4.1 ks; the main exception is LSS 1275, for which the telemetry saturation produced by the high level of the instrumental background reduced the exposure time to only 1.1 ks. We used version 4.5 of the \textit{CIAO} (Chandra Interactive Analysis of Observations)\footnote{http://asc.harvard.edu/ciao/} pipeline to reduce and analyse the data. For each source we reprocessed the data and accumulated the HRC image of the observed field-of-view. Then we extracted the source counts within a circular region with 2 arcsec radius, centred on the optical position of the star. For the background evaluation we extracted counts from an annular region around the source, with an inner and outer radius of 10 and 20 arcsec, respectively. We used the \texttt{ChaRT/MARX}\footnote{http://asc.harvard.edu/chart/runchart.html} package to reproduce the photon distribution on the focal plane due to the telescope \textit{Point Spread Function} (PSF) and to evaluate which PSF fraction is covered by the two extraction regions. Then we used the \textit{CIAO} task \textit{eprates} to measure the background- and PSF- corrected source count rates (CR) and their uncertainties.

\begin{table}[htbp]
\caption{Main parameters of the observations performed by \Chandra.}\label{observations}
\begin{center}
\begin{tabular}{cccc} \hline \hline
Target name		& ObsID	& Start Time		& Livetime (s)	\\ \hline
BD+37$^\circ$ 1977	& 14546	& 2013-09-29T01:56:21	& 4101		\\
BD+75$^\circ$ 325	& 14547	& 2013-04-21T19:29:14	& 3763		\\
BD+25$^\circ$ 4655	& 14548	& 2013-05-17T03:37:10	& 4139		\\
BD-22$^\circ$ 3804	& 14549	& 2013-05-22T05:50:52	& 4153		\\
BD+39$^\circ$ 3226	& 14550	& 2013-06-09T09:30:14	& 4149		\\
BD-03$^\circ$ 2179	& 14551	& 2013-09-07T22:53:24	& 4147		\\
BD+28$^\circ$ 4211	& 14552	& 2013-05-17T02:15:40	& 4130		\\
CD-31 4800		& 14553	& 2013-04-13T23:32:30	& 3932		\\
BD+48$^\circ$ 1777	& 14554	& 2013-10-10T19:16:39	& 4101		\\
LS V +22 38		& 14555	& 2013-09-07T21:14:34	& 4112		\\
LS IV -12 1		& 14556	& 2013-07-12T03:54:51	& 4153		\\
Feige 34		& 14557	& 2013-10-10T12:37:52	& 4148		\\
LS I +63 198		& 14558	& 2013-03-11T21:21:55	& 4147		\\
LSS 1275		& 14559	& 2013-03-17T05:51:29	& 1107		\\
LSE 153			& 14560	& 2013-07-26T23:54:40	& 4100		\\
LSE 21			& 14561	& 2013-09-02T08:57:12	& 4143		\\
LSE 263			& 14562	& 2013-03-03T17:25:37	& 3762		\\
BD+18$^\circ$ 2647	& 14563	& 2013-03-20T12:43:09	& 4096		\\
LS IV +10 9		& 14564	& 2013-07-16T19:28:06	& 4101		\\ \hline
\end{tabular}
\end{center}
\end{table}

For the sources detected at a confidence level higher than 3 $\sigma$, in Table~\ref{targets} we report the CR value; in the other cases we report the 3-$\sigma$ upper limit. We obtained a robust detection (\textit{signal-to-noise} ratio S/N $\simeq$ 4) only for \BDtwo\ and Feige 34; in addition, a marginal detection (S/N $\simeq$ 3) can be claimed for \BDthree. The CR of all other sources is consistent with 0 at a 3-$\sigma$ c.l. The detected CRs, or their upper limits for the undetected sources, are at level of a few 10$^{-3}$ c s$^{-1}$; the only exception is LSS 1275, for which the higher upper-limit value ($\simeq 10^{-2}$ c s$^{-1}$) is due to the larger CR uncertainty resulting from the shorter observing time. Since the HRC-I camera has essentially no energy resolution, no spectral study can be performed. To estimate the flux corresponding to the measured count rate or to its upper limit, we considered two different types of spectrum as representative of two physical origins of the source emission:
\begin{itemize}
\item for the hypothesis of emission from the subdwarf itself, we considered the spectrum observed for \HD\ during its eclipse phase, that is, a power law with photon index $\Gamma$ = 1.9 \citep{Mereghetti+13};
\item for the hypothesis that emission is caused by accretion onto a compact companion, we considered the soft spectrum observed for \HD\ out of its eclipse phase, that is, the sum of a blackbody component with kT = 30 eV, which provides the bulk of the source flux, and of a power-law component with $\Gamma$ = 2, which dominates above 0.5 keV \citep{Mereghetti+13}.
\end{itemize}
In both cases we used the \textit{WebPIMMs}\footnote{http://heasarc.gsfc.nasa.gov/Tools/w3pimms.html} tool to evaluate the CR-to-flux conversion factor (in the energy range 0.2-10 keV). To convert the flux into luminosity, we assumed a reasonable distance for the two stars without any distance estimate (LS V +22 38 and LS I +63 198) by comparing them with sources characterized by similar values of magnitude and surface effective temperature. We note that the source fluxes corresponding to the detected count rates (or to their upper limits for the undetected sources) are a few $10^{-14}$ erg cm$^{-2}$ s$^{-1}$, which is very near the sensitivity for which the observation campaign was originally planned.

\section{Discussion}\label{sec:4}

Our \Chandra\ survey resulted in the discovery of X-ray emission from three sdO stars. The brightest new source is \BDtwo, a luminous (log($g$) = 4) and He-rich sdO star \citep{JefferyHamann10} similar to \HD\ and \BDone, the only sdO stars previously known as X-ray sources. The other newly detected subdwarfs, Feige 34 and \BDthree, are instead compact (log($g$) $>$ 6) He-poor stars \citep{Thejll+91,ZaninWeinberger97}. The upper limits derived for the X-ray luminosity of most\footnote{The only exception is LSS 1275, which is characterized by a large uncertainty on both its flux and distance.} undetected sources (L$_X$$\lsim 10^{30}$ erg s$^{-1}$) are more constraining than those previously available and make accreting compact companions unlikely. The corresponding values of the X-ray/bolometric luminosity ratio agree with those observed for main-sequence, giant, and supergiant O-type stars \citep{Naze09}, which means that the undetected sdO might have a similar intrinsic X-ray emission as well.

The large distance of \BDtwo\ (d = 2.6 kpc) implies a luminosity $L_{\rm X} > 10^{31}$ erg s$^{-1}$, significantly higher than that of the other two new sources. This luminosity is similar to those estimated for \BDone\ and for the out-of-eclipse phase of \HD, which suggests that \BDtwo\ might also have an accreting compact companion. On the other hand, the bolometric luminosity of \BDtwo\ is higher than that of the other stars: $L_{\rm bol} \simeq 10^{38}$ erg s$^{-1}$ \citep{JefferyHamann10}, which implies log($L_{\rm X}/L_{\rm bol}$) $\simeq$ --6.3. This is within the range of values (--6.7$\pm$0.5) obtained by \citet{Naze09} for the single O-type stars observed by \XMM; moreover, a luminosity ratio within the same range would be obtained for any value of T$_{\rm eff}$ higher than the estimated one. We also note that \BDtwo\ is an extreme helium star with a significant stellar wind. By fitting the P-Cygni and asymmetric profiles of C, N, and Si ultraviolet resonance lines, which were obtained with high-resolution ultraviolet and optical spectra, \citet{JefferyHamann10} estimated a mass-loss rate $\dot M \simeq 10^{-8.2} M_{\odot}$ yr$^{-1}$ and a terminal wind velocity $v_{\infty} \simeq$ 2,000 km s$^{-1}$. Therefore, it is also plausible  that the observed X-ray flux is  produced by the wind internal shocks of the star itself. The comparison of the infrared flux of \BDtwo\ with that estimated on the basis of its T$_{\rm eff}$ indicates an excess at a low (2-$\sigma$) confidence level \citep{Ulla+98}; therefore, it is highly unlikely that a normal companion contributes to the X-ray flux.

The X-ray luminosity of Feige 34 is $\sim 10^{30}$ erg s$^{-1}$, which is at least one order of magnitude lower than that of \BDtwo. Compared with this, Feige 34 is significantly hotter (T$_{\rm eff} \simeq$ 70 kK) and has a much higher surface gravity (log($g$) = 7.3); on the other hand, for this star \citet{Thejll+95} estimated a similar mass-loss rate ($\dot M \simeq 10^{-7.5} M_{\odot}$ yr$^{-1}$). Its bolometric luminosity is relatively low (log($L_{\rm bol}/L_{\odot}$) = 2.6) and implies a bolometric/X-ray ratio log($L_{\rm X, 1}/L_{\rm bol}$) = --6.2: this agrees with the typical values obtained for normal O-type stars, unless the actual value of T$_{\rm eff}$ is significantly lower than the estimated one. These results suggest that the observed X-ray flux is generated by the internal shocks in the star wind, although the possibility of a low accretion rate onto a compact companion cannot be rejected. We finally note that Feige 34 shows an infrared excess at a 3-$\sigma$ level, which can be attributed to a companion star of spectral type M2 \citep{Thejll+95,Ulla+98}. Considering the optical luminosity expected for this type of stars, the observed X-ray flux implies --3 $\lsim$ log($f_{\rm X}/f_{\rm V}$) $\lsim$ --2. Since this value is in the range expected for M stars \citep{Krautter+99}, it would be important to assess whether there really is a late-type companion that might account for (or contribute to) the observed X-ray flux.

The small and well-constrained distance of \BDthree, together with the low detected count rate implies a very low luminosity ($\sim 10^{28}$ erg s$^{-1}$). It is about two orders of magnitude below that of Feige 34, although the two stars are rather similar: \BDthree\ is also hot (T$_{\rm eff}$ = 82 kK), has a high surface gravity (log($g$) = 6.2), and a relatively low bolometric luminosity (log($L_{\rm bol}/L_{\odot} \simeq$ 2). Its bolometric/X-ray ratio is low (log($L_{\rm X, 1}/L_{\rm bol}$) = --7.1), at the lower end of the range valid for O-type stars: this result supports  an intrinsic origin of the X-ray emission, although no evidence of significant stellar wind has been reported for \BDthree.  We also note that this star has a faint companion star at an angular separation of 2.8 arcsec \citep{MasseyGronwall90}; but the spatial resolution of the \Chandra\ images is high enough to exclude any contribution of this star to the detected X-ray flux.

%For all the remaining sources, we obtained no detection; we can only estimate that the upper limit on their X-ray luminosity is $\lsim 10^{30}$ erg s$^{-1}$, a rather low value which does not favour the possibility of matter accretion onto a compact companion\footnote{The only exception is LSS 1275, which is characterized by a large uncertainty on both its flux and distance.}. In all cases the corresponding values of the X-ray/bolometric luminosity ratio are in agreement with those observed for the main sequence, giant and supergiant O-type stars \citep{Naze09}. These results do not contradict the hypothesis that also the undetected sdO stars of our sample are characterized by a similar intrinsic X-ray emission; however, for those stars characterized by a significant excess in their infrared flux (see Table~\ref{targets}) the contribution of a possible companion star cannot be ignored. In summary, the performed \Chandra\ observations provided a detection of three sdO stars previously unknown as X-ray sources, but they do not allow us to provide any description about the properties and origin of the detected emission. To this aim, specific observations shall be performed.

\section{Conclusions}\label{sec:5}

We have carried out the first systematic X-ray survey of all the sdO stars brighter than V=12 using the \Chandra\ satellite. The upper limits for the 16 undetected sources do not exclude that the average relation between X-ray and bolometric luminosity derived for more luminous O stars also extends to the subdwarfs class. Thanks to the detection of \BDtwo, Feige 34, and possibly \BDthree, we more than doubled the number of sdO stars seen in the X-ray band. \BDtwo\ belongs to the subclass of luminous He-rich sdO, like the only two previously known X-ray emitting sdO, and, like these, it might have a compact companion. Feige 34 and \BDthree\ are instead compact sdO stars, the first of this kind to be detected in the X-ray band. Deeper X-ray observations are required to assess whether these three sdO stars contain compact accreting objects, or if their X-ray emission is intrinsic and due to shock-heating processes in the wind as is typical of more luminous early-type stars. In addition, more optical/IR studies are also required to exclude the presence of late-type companions that might contribute to the observed X-ray flux.

\begin{acknowledgements}
The scientific results reported in this article are based on observations made by the Chandra X-ray Observatory. This research has made use of software provided by the Chandra X-ray Center (CXC) in the application package CIAO. We acknowledge financial contributions by the Italian Space Agency through ASI/INAF agreement I/032/10/0. \end{acknowledgements}

\begin{landscape}
\begin{table}[htbp]
\caption{Main parameters of the observed targets. The bibliographic reference of the reported values is given in parenthesis.}\label{targets}
\begin{scriptsize}
\begin{center}
\begin{tabular}{ccccccccccccc} \hline \hline
(a)			& (b)				& (c)		& (d)			& (e)				& (f)		& (g)			& (h)				& (i)			& (j)						& (k)			& (l)				& (m)					\\
Name			& d				& V		& T$_{\rm eff}$		& log($g$)			& BC		& log($L_{\rm bol}$)	& CR    			& $f_{\rm X, 1}$	& $f_{\rm X, 2}$				& $L_{\rm X, 1}$    	& $L_{\rm X, 2}$		& log($\frac{L_{\rm X, 1}}{L_{\rm bol}}$)	\\
-			& (pc)				& (mag)		& (kK)			& (cm s$^{-2}$)			& (mag)		& ($L_{\odot}$)		& ($\times 10^{-3}$ c s$^{-1}$)	& \multicolumn{2}{c}{($\times 10^{-14}$ erg cm$^{-2}$ s$^{-1}$)}	& \multicolumn{2}{c}{($\times 10^{29}$ erg s$^{-1}$)}	& 					\\ \hline
BD+75$^\circ$ 325	& 136$^{+20}_{-16}$ (3)		& 9.55 (4)	& 55 (9)		& 5.5 (9)			& --4.76	& 2.21			& $<$ 1.6			& $<$ 2.7		& $<$ 1.2					& $<$ 0.6		& $<$ 0.3			& $<$ --7.05				\\
BD+25$^\circ$ 4655$^{*}$& 110$^{+16}_{-13}$ (3)		& 9.69 (2)	& 41.7 (2)		& 6.7 (2)			& --3.94	& 1.64			& $<$ 2.2			& $<$ 3.6		& $<$ 1.7					& $<$ 0.5		& $<$ 0.2			& $<$ --6.53				\\
BD--22$^\circ$ 3804	& 185$^{+53}_{-34}$ (3)		& 10.03 (2)	& 42.5 (10)		& 5.25 (10)			& --4.00	& 1.98			& $<$ 1.5			& $<$ 2.4		& $<$ 1.1					& $<$ 0.9		& $<$ 0.4			& $<$ --6.59				\\
BD+37$^\circ$ 1977$^{*}$& 2600 (1)			& 10.15 (19)	& 48 (1)		& 4 (1)				& --4.36	& 4.37			& 3.6$^{+1.1}_{-0.9}$		& 5.9$^{+1.8}_{-1.5}$	& 2.7$^{+0.8}_{-0.7}$				& 450$^{+140}_{-110}$	& 210$^{+60}_{-50}$		& --6.30$\pm$0.12			\\
BD+39$^\circ$ 3226	& 235$^{+88}_{-50}$ (3)		& 10.18 (2)	& 45$\pm$5 (11)		& 5.5$\pm$0.5 (11)		& --4.15	& 2.19			& $<$ 1.4			& $<$ 2.4		& $<$ 1.1					& $<$ 1.5		& $<$ 0.7			& $<$ --6.60				\\
BD--03$^\circ$ 2179	& 631$_{-318}$ (3)		& 10.33 (2)	& 62$\pm$5 (5)		& 4.5$\pm$0.5 (5)		& --5.12	& 3.38			& $<$ 3.2			& $<$ 5.2		& $<$ 2.4					& $<$ 23		& $<$ 11			& $<$ --6.59				\\
BD+28$^\circ$ 4211	& 92$^{+14}_{-11}$ (3)		& 10.51 (4)	& 82$\pm$5 (12)		& 6.2$^{+0.3}_{-0.1}$ (12)	& --5.95	& 1.96			& 1.8$^{+0.8}_{-0.7}$		& 2.9$^{+1.3}_{-1.1}$	& 1.3$^{+0.6}_{-0.5}$				& 0.3$\pm$0.1		& 0.13$^{+0.06}_{-0.05}$	& --7.10$^{+0.16}_{-0.21}$		\\
CD--31 4800		& 132$^{+34}_{-10}$ (3)		& 10.52 (2)	& 44$\pm$1 (10)		& 5.4$\pm$0.3 (10)		& --4.10	& 1.53			& $<$ 1.5			& $<$ 2.5		& $<$ 1.2					& $<$ 0.5		& $<$ 0.2			& $<$ --6.42				\\
BD+48$^\circ$ 1777$^{*}$& 163$^{+65}_{-36}$ (3)		& 10.74 (5)	& 40 (5)		& -				& --3.82	& 1.52			& $<$ 2.3			& $<$ 3.8		& $<$ 1.7					& $<$ 1.1		& $<$ 0.5			& $<$ --6.04				\\
LS V +22 38$^{*}$	& 180$^{*}$			& 10.93 (6)	& 40 (12)		& -				& --3.82	& 1.53			& $<$ 2.3			& $<$ 3.8		& $<$ 1.7					& $<$ 1.4		& $<$ 0.6			& $<$ --5.96				\\
LS IV --12 1		& 400$\pm$150 (7)		& 11.16 (2)	& 60 (2)		& 4.5 (2)			& --5.02	& 2.61			& $<$ 1.4			& $<$ 2.4		& $<$ 1.1					& $<$ 4.3		& $<$ 2.0			& $<$ --6.56				\\
Feige 34$^{*}$		& 325$^{+540}_{-125}$ (3)	& 11.18 (4)	& 70$\pm$10 (5)		& 7.3$\pm$0.4 (5)		& --5.48	& 2.60			& 4.6$^{+1.2}_{-1.0}$		& 7.5$^{+2.0}_{-1.6}$	& 3.4$^{+0.9}_{-0.7}$				& 17.0$^{+4.4}_{-3.7}$	& 7.8$^{+2.0}_{-1.7}$		& --6.23$^{+0.10}_{-0.11}$		\\
LSE 153			& 250$\pm$100 (7)		& 11.36 (2)	& 70.0$\pm$1.5 (2)	& 4.75$\pm$0.15 (2)		& --5.48	& 2.30			& $<$ 3.0			& $<$ 5.0		& $<$ 2.3					& $<$ 3.6		& $<$ 1.6			& $<$ --6.34				\\
LSS 1275		& $<$ 1000 (8)			& 11.37 (2)	& 75$\pm$4 (2)		& 5.0$\pm$0.2 (2)		& --5.69	& 3.59			& $<$ 10.7			& $<$ 17.5		& $<$ 8.0					& $<$ 200		& $<$ 90			& $<$ --5.87				\\
LSE 263			& 250$\pm$100 (7)		& 11.55 (19)	& 70.0$\pm$2.5 (2)	& 4.90$\pm$0.25 (2)		& --5.48	& 2.23			& $<$ 3.8			& $<$ 6.2		& $<$ 2.8					& $<$ 4.4		& $<$ 2.0			& $<$ --6.17				\\
BD+18$^\circ$ 2647	& 275$^{+450}_{-105}$ (3)	& 11.63 (19)	& 75$\pm$5 (10)		& 5.2$\pm$0.2 (10)		& --5.69	& 2.36			& $<$ 2.3			& $<$ 3.8		& $<$ 1.7					& $<$ 3.2		& $<$ 1.5			& $<$ --6.44				\\
LSE 21			& 50 (7)			& 11.64 (19)	& 110$\pm$10 (2)	& -				& --6.82	& 1.33			& $<$ 1.5			& $<$ 2.4		& $<$ 1.1					& $<$ 0.07		& $<$ 0.03			& $<$ --7.08				\\
LS IV +10 9$^{*}$	& 230$\pm$100 (7)		& 12.05 (2)	& 45 (13)		& 5.6 (13)			& --4.17	& 1.43			& $<$ 1.5			& $<$ 2.4		& $<$ 1.1					& $<$ 1.4		& $<$ 0.7			& $<$ --5.86				\\
LS I +63 198$^{*}$	& 200$^{*}$			& 12.80$^{*}$ (20)	& 34$\pm$7 (2)		& 5.4$\pm$0.3 (2)		& --3.33	& 0.67			& $<$ 1.8			& $<$ 3.0		& $<$ 1.4					& $<$ 1.4		& $<$ 0.6			& $<$ --5.13				\\ \hline
\HD\			& 650$\pm$100 (14)		& 8.29 (4)	& 46.5 (15)		& 4.35 (15)			& --4.27	& 4.15			& -				& 7.3$\pm$0.7$^{*}$ (16)& 66$\pm$1$^{*}$ (16)				& 35$\pm$3		& 315$\pm$5			& --7.21$\pm$0.05			\\
\BDone\			& 2000$^{+900}_{-600}$ (10)	& 10.01 (17)	& 48 (1)		& 4.00$\pm$0.25 (10)		& --4.36	& 4.40			& -				& -			& 2.6$\pm$0.3$^{*}$ (18)			& -			& 120$\pm$14			& -					\\ \hline \hline
\end{tabular}
\end{center}
\end{scriptsize}
\begin{small}
Key to table: (a) source name ($^{*}$ sources with an infrared flux excess detected at a confidence level higher than 2 $\sigma$ \citep{Thejll+95,Ulla+98}); (b) source distance ($^{*}$ assumed distance); (c) apparent visual magnitude ($^{*}$ LS I +63 198 was included in our sample based on the brighter magnitude provided by \citet{Ostensen06}); (d) surface effective temperature; (e) surface gravity; (f) Bolometric Correction; (g) bolometric luminosity; (h) detected count rate (or 3-$\sigma$ upper limit for undetected sources); (i) flux in the energy range 0.2-10 keV, based on the measured count rate or count-rate upper limit, assuming a power-law spectrum ($^{*}$ measured flux during source eclipse); (j) flux value in the energy range 0.2-10 keV, based on the measured count rate or count-rate upper limit, assuming a power-law plus blackbody spectrum ($^{*}$ measured flux, out of source eclipse for \HD); (k) estimated luminosity, based on $f_{\rm X, 1}$; (l) estimated luminosity, based on $f_{\rm X, 2}$; (m) estimated value of X-ray/bolometric luminosity ratio, based on $f_{\rm X, 1}$
\\
References:  1 - \citet{JefferyHamann10}; 2 - \citet{Ostensen06}; 3 - \citet{vanLeeuwen07}; 4 - \citet{LandoltUomoto07}; 5 - \citet{Thejll+95}; 6 - \citet{Hog+98}; 7 - \citet{SchonbernerDrilling84}; 8 - \citet{Rauch+91}; 9 - \citet{Lanz+97}; 10 - \citet{BauerHusfeld95}; 11 - \citet{RodriguezLopez+07}; 12 - \citet{Latour+13}; 13 - \citet{Ulla+98}; 14 - \citet{KudritzkiSimon78}; 15 -  \citet{Hamann10}; 16 - \citet{Mereghetti+13}; 17 - \citealt{Landolt73}; 18 - \citet{LaPalombara+12}; 19 - \citet{Hog+00}; 20 - \citet{Reed03}
\end{small}
\end{table}
\end{landscape}

\bibliographystyle{aa}
\bibliography{biblio}

\end{document}